\documentclass
[twocolumn,aps,prd,amsmath,showpacs,nofootinbib,floatfix]
{revtex4-1}

\usepackage[dvips]{graphicx}
\usepackage{bm}                      
\usepackage{mathptmx}                
\usepackage{dcolumn}                 

\def\bea{\begin{eqnarray}}
\def\eea{\end{eqnarray}}
\def\be{\begin{equation}}
\def\ee{\end{equation}}

\def\section#1{{\bf #1.---\hskip-1mm}} 

\begin{document}

\title{
Nucleon effective masses within the Brueckner-Hartree-Fock theory:\\
Impact on stellar neutrino emission}

\author{
M. Baldo, G. F. Burgio, H.-J. Schulze, and G. Taranto}

\affiliation{
INFN Sezione di Catania and
Dipartimento di Fisica e Astronomia,
Universit\'a di Catania, Via Santa Sofia 64, 95123 Catania, Italy}

\date{\today}

\begin{abstract}
We calculate the effective masses of neutrons and protons
in dense nuclear matter
within the microscopic Brueckner-Hartree-Fock many-body theory
and study the impact on the neutrino emissivity processes of neutron stars.
We compare results based on different nucleon-nucleon potentials
and nuclear three-body forces.
Useful parametrizations of the numerical results are given.
We find substantial in-medium suppression of the emissivities,
strongly dependent on the interactions.
\end{abstract}

\pacs{
26.60.-c,  
26.60.Kp,  
97.10.Cv.  
}

\maketitle

\section{Introduction}
With the commissioning of increasingly sophisticated instruments,
more and more details of the very faint signals emitted by neutron stars (NS)
can be quantitatively monitored.
This will allow in the near future an ever increasing accuracy to constrain
the theoretical ideas for the ultra-dense matter that composes these objects.

One important tool of analysis is the temperature-vs.-age cooling diagram,
in which currently a few observed NS are located.
NS cooling is over a vast domain of time ($10^{-10}$--$10^5$ yr)
dominated by neutrino emission due to several microscopic processes \cite{rep}.
The theoretical analysis of these reactions requires,
apart from the elementary matrix elements,
the knowledge of the density of states of the relevant reaction partners
and thus the nucleon effective masses.

The present report is focused on the problem
of the theoretical determination of this important input information
and reports nucleon effective masses in dense nuclear matter
obtained within the Brueckner-Hartree-Fock (BHF)
theoretical many-body approach. 
We study the dependence on the underlying basic two-nucleon and
three-nucleon interactions and
provide useful parametrizations of the numerical results.
Finally some estimates of the related in-medium modification
of the various neutrino emission rates in NS matter will be given.
We begin with a short review of the BHF formalism and the relevant
neutrino emission processes, before presenting our numerical results.

\section{The Brueckner-Hartree-Fock approach}
\label{s:eos}
Empirical properties of infinite nuclear matter can be calculated using
many different theoretical approaches.
In this paper we concentrate on the non-relativistic BHF method,
which is based on a linked-cluster
expansion of the energy per nucleon of nuclear matter \cite{bhf,book,bb}.
The basic ingredient in this many-body approach is the reaction
matrix $G$, which is the solution of the Bethe-Goldstone equation
\be
 G[\rho;\omega] = V + \sum_{k_a k_b} V {{|k_a k_b\rangle  Q  \langle k_a k_b|}
 \over {\omega - e(k_a) - e(k_b) }} G[\rho;\omega] \:,
\label{e:g}
\ee
where $V$ is the bare nucleon-nucleon (NN) interaction,
$\rho$ is the nucleon number density,
and $\omega$ the starting energy.
The single-particle (s.p.) energy
\be
 e(k) = e(k;\rho) = {k^2\over 2m} + U(k;\rho)
\label{e:en}
\ee
and the Pauli operator $Q$ determine the propagation
of intermediate baryon pairs.
The BHF approximation for the s.p.~potential
$U(k;\rho)$ using the {\em continuous choice} is
\be
 U(k;\rho) = {\rm Re} \sum _{k'\leq k_F}
 \big\langle k k'\big| G[\rho; e(k)+e(k')] \big| k k'\big\rangle_a \:,
\ee
and the energy per nucleon is then given by
\be
 {E \over A} =
 {3\over5}{k_F^2\over 2m} + {1\over{2\rho}} \sum_{k,k'\leq k_F}
 \big\langle k k'\big| G[\rho; e(k)+e(k')] \big|k k'\big\rangle_a \:,
\ee
where the subscript $a$ indicates antisymmetrization of the matrix element.
In this scheme, the only input quantity needed is the bare NN interaction
$V$ in the Bethe-Goldstone equation~(\ref{e:g}).

The nuclear EOS can be calculated with good accuracy in this two hole-line
approximation with the continuous choice for the s.p.~potential,
since the results in this scheme are quite close to those
which include also the three hole-line contribution \cite{song}.
The dependence on the NN interaction,
also within other many-body approaches,
has recently been systematically investigated in Refs.~\cite{satu,polls}.

\begin{figure}[t]
\vspace{3mm}
\includegraphics[angle=0,scale=0.52,clip]{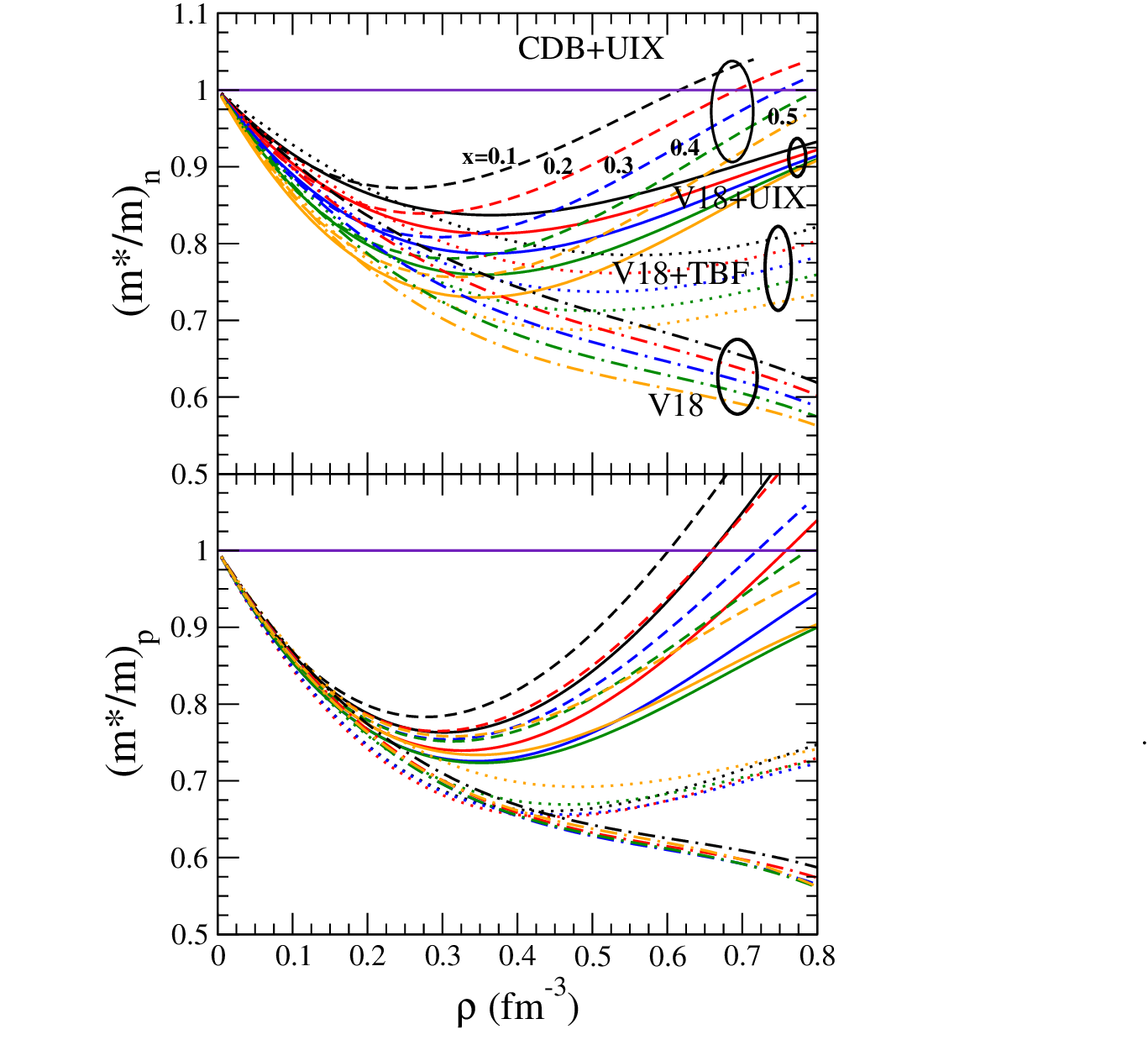}
\vspace{-8mm}
\caption{(Color online)
The neutron/proton (top/bottom panel)
effective mass is displayed vs.~the nucleon density
for several values of the proton fraction
$x=0.1, 0.2, 0.3, 0.4, 0.5$.
Results are plotted for different choices of two- and three-body forces,
as discussed in the text.}
\label{f:meff}
\end{figure}

However, it is commonly known that non-relativistic calculations,
based on purely two-body interactions,
fail to reproduce the correct saturation point of symmetric nuclear matter,
which requires the introduction of three-body forces (TBF).
In our approach, following Ref.~\cite{tbfmic},
the TBF is reduced to a density-dependent
two-body force by averaging over the position of the third particle,
assuming that the
probability of having two particles at a given distance is reduced
according to the two-body correlation function \cite{bbb,zhou}.
More precisely, in the current procedure any exchange diagrams involving
the in-medium particle are neglected,
but the proper spin-isospin correlations in the relative
$^1S_0$ and $^3S_1$ states
are maintained via the corresponding defect functions.
For more details we refer to \cite{tbfmic,zhli} and references therein.
For the moment a completely consistent inclusion of TBF in the BHF formalism
is not yet achieved, although there is some recent progress \cite{artur}.

Following this procedure,
we will illustrate results for two different approaches to the
TBF, i.e., a phenomenological and a microscopic one.
The phenomenological approach is based on the so-called Urbana model \cite{uix}.
The two parameters contained in this TBF have been fine-tuned
to get an optimal saturation point \cite{alad}
for the Argonne $V_{18}$ \cite{v18}
or the CD-Bonn potential \cite{cdb} that we use in the following.

The connection between two-body and three-body forces within the
meson-nucleon theory of the nuclear interaction is extensively discussed and
developed in references \cite{tbfmic,zhli}.
At present the theoretical status of microscopically derived TBF
is still quite rudimentary,
however a tentative approach has been proposed using the same meson-exchange
parameters as the underlying NN potential.
Results have been obtained with the
Argonne $V_{18}$, the Bonn B, and the Nijmegen 93
potentials \cite{zhli,nstbf}.

The nucleon effective mass $m^*$ describes the nonlocality of the s.p.~potential
felt by a nucleon propagating in the nuclear medium.
It is of great interest since it is closely related
to many nuclear phenomena such as the
dynamics of heavy-ion collisions at intermediate and high energies,
the damping of nuclear excitations and giant resonances,
and the adiabatic temperature of collapsing stellar matter.
The momentum-dependent effective mass is defined in terms of the s.p.~energy,
\be
 \frac{m^*(k)}{m} = \frac{k}{m} \left[ \frac{d e(k)}{dk} \right]^{-1}
\ee
and clearly arises from both the momentum and energy dependence
of the microscopic s.p.~potential \cite{mbhf}.
For the applications we consider the effective
mass taken at the Fermi surface $k_{F_{n,p}}$.

In this work we analyze the effective mass obtained in the lowest-order
BHF approximation discussed above.
It is well know that including second-order ``rearrangement'' contributions to
the s.p.~potential increases the theoretical $m^*$ values \cite{mfp,mbhf},
see also recent perturbative calculations \cite{chiral}.
However, the inclusion of the rearrangement term would require also the
re-examination of the EOS,
since the three hole-line contribution is altered by the modification
of the s.p.~potential,
and it could be not any more negligible \cite{book,bb}.
We defer the analysis of this point to later work
and concentrate here rather on the dependence of the results
with respect to the choice of the two-body and three-body forces.

NS matter is composed of asymmetric nuclear matter,
where the effective mass depends both on the nucleon density and on the
proton fraction $x = \rho_p/\rho$.
The BHF neutron and proton effective masses
in asymmetric matter are displayed in Fig.~\ref{f:meff}
as a function of the nucleon density for several values of the proton fraction.
Different choices of the NN potential and TBF are compared,
namely, we display results for the Argonne $V_{18}$ potential
without TBF (V18, dash-dotted lines),
with microscopic TBF (V18+TBF, dotted lines),
with phenomenological Urbana TBF (V18+UIX, solid lines),
and for the CD Bonn potential plus Urbana TBF (CDB+UIX, dashed lines).
We see that without TBF the values of the effective masses decrease
with increasing nucleon density,
whereas the inclusion of TBF causes an increase of the values
at densities above 0.3--0.4 fm$^{-3}$
for both protons and neutrons and all considered models.
This is due to the repulsive character of the TBF at high density.
There is evidently a strong dependence on the chosen set of interactions,
which reflects in particular
the current theoretical uncertainty regarding nuclear TBF at high density.

For easy implementation in astrophysical applications,
we provide polynomial fits of the effective masses
(valid for $\rho\leq0.8\;\text{fm}^{-3}$),
\bea
 {m^* \over m}(\rho,x) = 1
 &-& (a_1 + b_1x + c_1x^2) \rho
\nonumber\\[-1.5mm]
 &+& (a_2 + b_2x + c_2x^2) \rho^2
\nonumber\\
 &-& (a_3 + b_3x + c_3x^2) \rho^3 \:,
\label{e:fit}
\eea
whose parameters are reported in Table~\ref{t:fit}.

\setlength{\extrarowheight}{2pt}
\squeezetable
\begin{table}[t]
\caption{
Parameters of the polynomial fits, Eq.~(\ref{e:fit}),
for the neutron and proton effective masses,
obtained with different interactions.
}
\begin{ruledtabular}
\begin{tabular}{ld|ddddddddd}
&&\multicolumn{1}{c}{$a_1$}&\multicolumn{1}{c}{$b_1$}&\multicolumn{1}{c}{$c_1$}
&\multicolumn{1}{c}{$a_2$}&\multicolumn{1}{c}{$b_2$}&\multicolumn{1}{c}{$c_2$}
&\multicolumn{1}{c}{$a_3$}&\multicolumn{1}{c}{$b_3$}&\multicolumn{1}{c}{$c_3$}
\\
\hline
 V18     &p& 1.45 & 0.85 &-0.92 & 2.10 & 1.26 &-0.44 & 1.13 & 0.65 & 0.42 \\
         &n& 0.96 & 0.92 & 0.59 & 1.20 & 1.38 & 1.64 & 0.71 & 0.65 & 0.98 \\
 V18+TBF &p& 1.67 & 0.99 &-2.47 & 2.7 & 1.18 &-3.75 & 1.14 & 0.88 &-2.4 \\
         &n& 0.61 & 1.55 & 0.91 & 0.42 & 2.01 & 4.77 &-0.17 & 0.58 & 4.44 \\
 V18+UIX &p& 1.56 & 1.31 &-1.89 & 3.17 & 1.26 &-1.56 & 0.79 & 3.78 &-3.81 \\
         &n& 0.88 & 1.21 & 1.07 & 1.64 & 2.06 & 2.87 & 0.78 & 0.98 & 1.62 \\
 CDB+UIX &p& 1.53 & 0.80 &-1.04 & 3.05 & 1.06 &-1.44 & 0.43 & 4.04 &-4.42 \\
         &n& 0.95 & 1.17 & 0.42 & 2.44 & 1.27 &-0.05 & 1.30 & 0.55 &-1.63 \\
\end{tabular}
\end{ruledtabular}
\label{t:fit}
\end{table}

\section{Neutrino emissivities}
In this section we briefly recall the main neutrino emission mechanisms
in NS and the relevance of the nucleon effective masses,
following closely the detailed treatment given in Ref.~\cite{rep}.
Only the rates for the non-superfluid scenarios will be given,
for which the dependence on the effective masses is via the general factor
\be
 M_{ij} \equiv  \left( \frac{\rho_p}{\rho_0} \right)^{1/3} \!\!
 \widetilde{M}_{ij} \ , \quad
 \widetilde{M}_{ij} \equiv
 \left(\frac{m_n^*}{m_n}\right)^i \left(\frac{m_p^*}{m_p}\right)^j \:.
\label{e:m}
\ee
In the presence of superfluidity the dependence is highly nontrivial
and requires detailed calculations \cite{rep}.
In the following all emissivities $Q$ are given in units of
$\rm erg~ cm^{-3}~ s^{-1}$.

In the absence of pairing
three main mechanisms are usually taken into account,
which are the direct Urca (DU), the modified Urca (MU),
and the NN bremsstrahlung (BNN) processes.
By far the most efficient mechanism of NS cooling is the DU process,
for which the derivation of the emissivity under the condition of $\beta$
equilibrium is based on the $\beta$-decay theory \cite{lat91}.
The result for $npe$ NS matter is given by
\be
 Q^{(DU)} \approx
 4.0\times10^{27} M_{11} T_9^6 \mathop\Theta(k_{F_p} + k_{F_e} - k_{F_n}) \:,
\label{e:du}
\ee
where $T_9$ is the temperature in units of $10^9$K.
If muons are present,
then the equivalent DU process may also become possible,
in which case the neutrino emissivity is increased by a factor of 2.

The emissivities of the MU
processes in the neutron  and proton branches \cite{fm79, yl95} are given respectively by
\bea
 Q^{(Mn)} &\approx& 8.1\times10^{21} M_{31} T_9^8 \alpha_n \beta_n \:,
 \label{e:mun}
\\
 Q^{(Mp)} &\approx& 8.1\times10^{21} M_{13} T_9^8 \alpha_p \beta_p
 \left( 1 - k_{F_e}/4k_{F_p} \right)
 \Theta_{Mp} \:,
\label{e:mup}
\eea
where the factor $\alpha_n$ ($\alpha_p$) takes into account the momentum transfer
dependence of the squared reaction matrix element of the neutron (proton)
branch under the Born approximation, and $\beta_n$ ($\beta_p$) includes the
non-Born corrections due to the NN interaction effects,
which are not described by the one-pion exchange \cite{rep}.
The currently adopted values are
$\alpha_p=\alpha_n=1.13$ and
$\beta_p=\beta_n=0.68$.
The main difference between the proton branch and the neutron branch
is the threshold character,
since the proton branch is allowed only if
$k_{F_n} < 3k_{F_p} + k_{F_e}$,
in which case $\Theta_{Mp}=1$.
If muons are present in the dense NS matter,
the equivalent MU processes become also possible.
Accordingly, several modifications should be included in
Eqs.~(\ref{e:mun},\ref{e:mup}),
as discussed in Ref.~\cite{rep}.

Following the discussion above, the
neutrino emissivity jumps directly from the value of the MU
process to that of the DU process.
Thus, the DU process appears in a step-like manner.
In the absence of the DU process, the standard neutrino luminosity
of the $npe$ matter is determined not only by the MU processes
but also by the BNN processes in NN collisions:
\be
 N + N \rightarrow N + N + \nu + \overline\nu \:.
\ee
These reactions proceed via weak neutral currents and produce neutrino pairs of
any flavor \cite{fm79,yl95}. In analogy with the MU process, the
emissivities depend on the employed model of NN interactions.
Contrary to the MU, an elementary act of the NN bremsstrahlung
does not change the composition of matter.
The BNN has evidently no thresholds associated with
momentum conservation and operates at any density in the uniform matter.
The neutrino emissivity of the BNN processes in $npe$ NS matter is
\bea
 Q^{(Bnn)} &\approx& 2.3 \times 10^{20} M_{40} T_9^8
 \alpha_{nn} \beta_{nn} {(\rho_n/\rho_p)^{1/3}} \:,
\label{e:nn}\\
 Q^{(Bnp)} &\approx& 4.5 \times 10^{20} M_{22} T_9^8
 \alpha_{np} \beta_{np} \:,
\label{e:np}\\
 Q^{(Bpp)} &\approx& 2.3 \times 10^{20} M_{04} T_9^8
 \alpha_{pp} \beta_{pp} \:.
\label{e:pp}
\eea
The dimensionless factors $\alpha_{NN}$ come from the estimates of the
squared matrix elements at $\rho = \rho_0$:
$\alpha_{nn}=0.59$, $\alpha_{np}=1.06$, $\alpha_{pp}=0.11$.
The correction factors $\beta_{NN}$ are taken as
$\beta_{nn}=0.56$, $\beta_{np}=0.66$, $\beta_{pp}=0.70$.
All three processes are of comparable intensity,
with $Q^{(Bpp)} < Q^{(Bpn)} < Q^{(Bnn)}$.

\section{Results for $\beta$-stable matter}
\label{s:res}
\begin{figure}[t]
\includegraphics[angle=0,scale=0.48,clip]{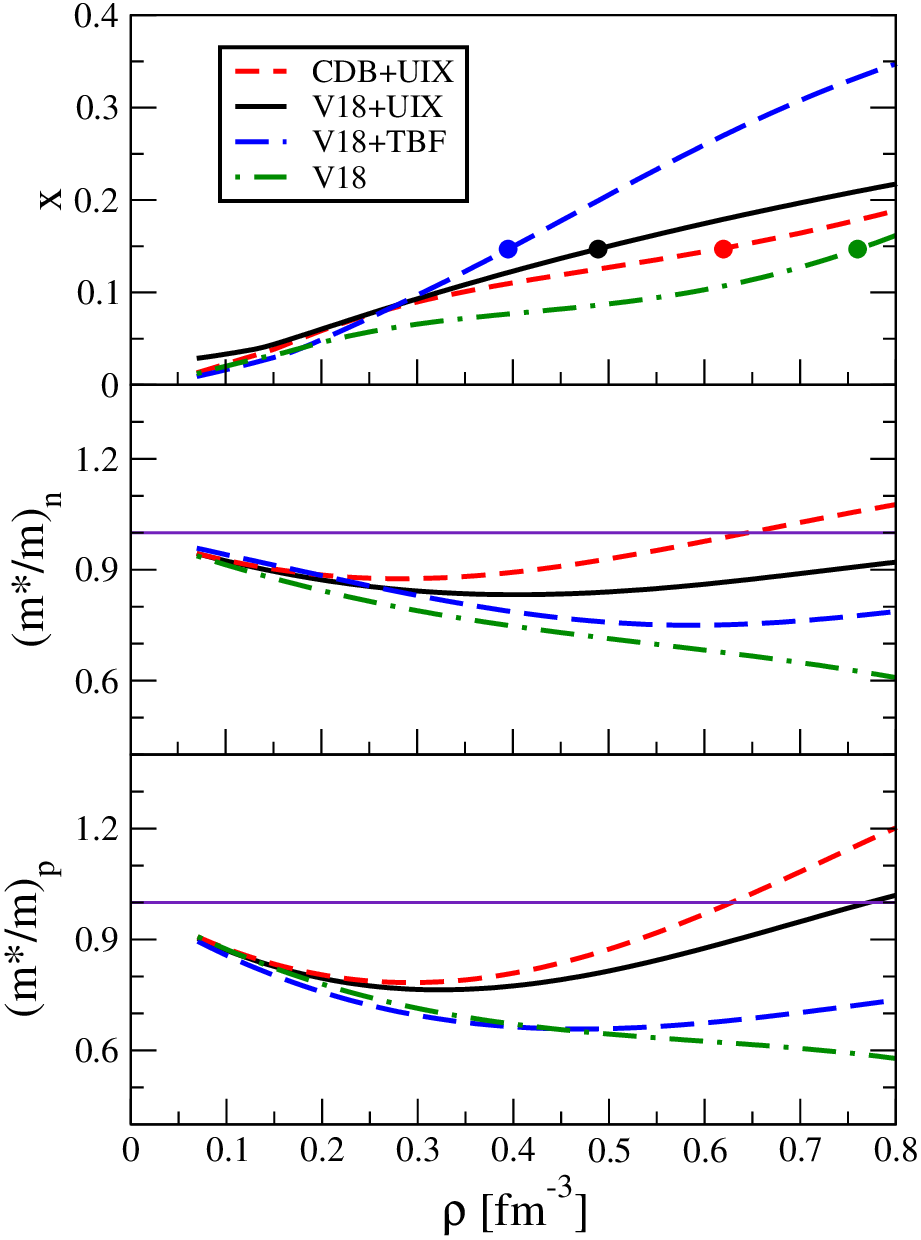}
\caption{(Color online)
Proton fraction (upper panel; the dots indicate the onset of the DU process)
and neutron/proton effective masses (central/lower panel)
in $\beta$-stable matter obtained with different interactions.
}
\label{f:mstable}
\end{figure}

For the treatment of NS matter we assume as usual
charge neutral,
$\rho_p = \rho_e + \rho_\mu$,
and $\beta$-stable,
$\mu_n - \mu_p = \mu_e = \mu_\mu$,
nuclear matter.
In Fig.~\ref{f:mstable} we show our results for this case,
obtained with the different combinations of
two- and three-body potentials introduced before.
In the upper panel the proton fraction
is displayed as a function of the nucleon density,
whereas the middle and lower panels show the neutron and proton
effective masses, respectively.

\begin{figure*}[t]
\includegraphics[scale=0.65,clip]{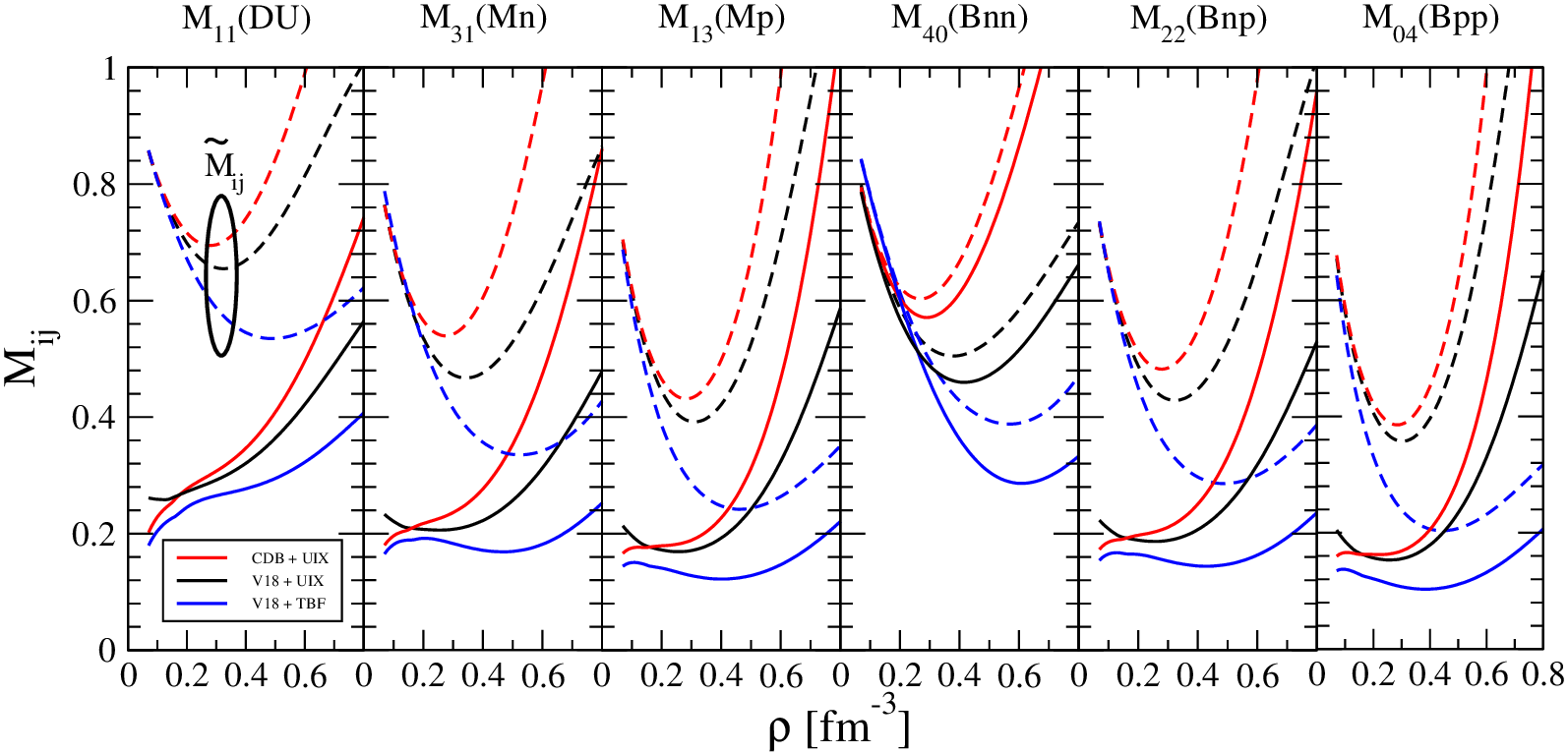}
\caption{(Color online)
Reduction factors $M_{ij}$ (solid lines)
and $\widetilde{M}_{ij}$ (dashed lines)
for the various cooling processes in $\beta$-stable matter
obtained with different interactions.
See text for details.
}
\label{f:urca}
\end{figure*}

We observe that the inclusion of TBF increases
the proton fraction \cite{zhou,nstbf,zuo}
due to the increased repulsion at large density,
leading to the onset of the DU process in all cases
(at different threshold densities indicated by markers).
The effective masses also start to increase at high density
due to the action of TBF,
but depend strongly on the interactions:
the V18+TBF model predicts the strongest
and the CDB+UIX the weakest medium effects.
Note that the value of the effective mass in $\beta$-stable matter
obtained with different interactions is a consequence,
apart from the differences shown in Fig.~\ref{f:meff},
also of the different proton fractions,
as shown in the upper panel of Fig.~\ref{f:mstable}.

Finally we combine the results shown in
the various panels of Fig.~\ref{f:mstable}
in order to obtain the reduction factors $M_{ij}$, $\widetilde{M}_{ij}$,
Eq.~(\ref{e:m}), for the different cooling processes.
Fig.~\ref{f:urca} displays the different factors
$M_{11}$ (for DU),
$M_{31}$, $M_{31}$ (for MU), and
$M_{40}$, $M_{22}$, $M_{04}$ (for BNN)
[solid curves]
and the corresponding $\widetilde{M}_{ij}$ factors
[dashed curves].
In line with the previous discussion,
one notes again the strong interaction dependence
of both the complete factors $M_{ij}$ and the
in-medium modification factors $\widetilde{M}_{ij}$.
The latter show generally
(apart from the CDB+UIX at high density)
a reduction of the emissivities due to the
general in-medium reduction of the effective masses.

\section{Conclusions}
\label{s:end}
We have computed nucleon effective masses in the BHF formalism
for dense nuclear matter, employing different combinations of
two-nucleon and three-nucleon forces.
Useful parametrizations of the numerical results were provided.
The relevant in-medium correction factors for several neutrino
emission processes in $\beta$-stable non-superfluid
neutron star matter have then been evaluated in a consistent manner.
We find in general in-medium suppression of the emissivities,
which however depends strongly on the employed interactions,
and reflect mainly the current lack of knowledge
regarding nuclear TBF at high density.
This emphasizes the need of performing and comparing
consistent calculations
with given sets of two-body and three-body interactions.



\end{document}